\documentclass[11pt,a4paper]{article}
\usepackage[T1]{fontenc}
\usepackage[utf8]{inputenc}
\usepackage{lmodern}
\usepackage[margin=25mm]{geometry}
\usepackage{microtype}
\usepackage{amsmath,amssymb}
\usepackage{graphicx}
\usepackage{booktabs,array,tabularx}
\usepackage{caption}
\usepackage{enumitem}
\usepackage[numbers,sort&compress]{natbib}
\usepackage{xurl}
\usepackage{fancyhdr}
\usepackage{placeins}
\usepackage{float}
\usepackage[hidelinks]{hyperref}

\newcolumntype{P}[1]{>{\raggedright\arraybackslash}p{#1}}
\newcolumntype{Y}{>{\raggedright\arraybackslash}X}
\renewcommand{\arraystretch}{1.15}
\setlist[itemize]{nosep,leftmargin=1.4em}

\begin{document}

\thispagestyle{plain}

\begin{center}
{\LARGE\bfseries A Quantum Chemistry Dataset of 120 Kilo Organic Molecules with 20 Million Conformations\par}
\vspace{12pt}
{\normalsize Weiqi Liu$^{1}$, Xi Ai$^{2}$, Zhijian Zhou$^{1}$, Junyi An$^{3}$, Chao Qu$^{1}$,\\[3pt]
Zhipeng Zhou$^{2}$, Yuan Cheng$^{1}$, Yinghui Xu$^{1}$, Fenglei Cao$^{3}$ and Yuan Qi$^{1,3}$\par}

\renewcommand{\thefootnote}{}
\footnotetext{\small$^{1}$Fudan University, Shanghai, China.
$^{2}$INFTech, Shanghai, China.
$^{3}$SAIS, Shanghai, China.
\par
Correspondence to: Fenglei Cao (caofenglei@sais.com.cn) and Yuan Qi (qiyuan@fudan.edu.cn).\par}
\end{center}

\section{Abstract}
Quantum-chemical datasets of molecular geometries, electronic energies and atomic forces support the development of atomistic machine-learning models. Here we present QO2Mol, comprising 116,127 organic molecular fragments and more than 20 million three-dimensional conformations. The fragments were obtained by bond cleavage and capping of source compounds and contain H, C, N, O, F, P, S, Cl, Br and I. The geometries include optimized structures and configurations generated through torsional sampling and local bond-length and bond-angle perturbations. Electronic energies and atomic forces were calculated at the B3LYP/def2-SVP level of theory. Each conformational record also contains dipole moments, isotropic polarizabilities, electronic spatial extents and frequency-derived correction values. The dataset and its documented generation and validation procedures support molecular energy and force modelling and studies of geometry-dependent molecular properties.

\section{Background \& Summary}
Quantum-chemical reference data have played an important role in the development of atomistic models of molecular energies and forces~\cite{Behler2007,Smith2017,Hoja2021}. Calculated properties for specified atomic configurations provide a common basis for investigating molecular structure and for training and evaluating machine-learning models. Because molecular energies and forces vary with nuclear coordinates, collections comprising multiple geometries of each molecular structure also support studies of conformational changes and non-equilibrium configurations.

Established quantum-chemical datasets encompass a range of molecular systems, conformational sampling approaches and calculated properties. QM9 reports calculated properties for small organic molecules~\cite{Ramakrishnan2014}, and ANI-1 provides extensive sampling of non-equilibrium molecular geometries~\cite{Smith2017}. QM7-X combines equilibrium and non-equilibrium structures with a broad selection of quantum-mechanical properties~\cite{Hoja2021}. Further resources, including QMugs~\cite{Isert2022}, SPICE~\cite{Eastman2023} and the $\nabla$DFT and $\nabla^{2}$DFT collections~\cite{Khrabrov2022,Khrabrov2024}, contribute additional molecular systems, conformational sampling strategies and calculated properties. Together, these datasets have substantially advanced the availability of reference data for molecular modelling.

Here, we present QO2Mol, a quantum-chemical dataset comprising 116,127 organic molecular fragments and more than 20 million conformations. The fragments were prepared through bond cleavage and capping of source compounds and encompass ten elements: H, C, N, O, F, P, S, Cl, Br and I. The dataset includes optimized geometries and non-equilibrium configurations generated through torsional sampling and local bond-length and bond-angle perturbations. Electronic energies and atomic forces were calculated at the B3LYP/def2-SVP level of theory.

In addition to energies and forces, the released records contain geometry-dependent electronic descriptors, including dipole moments, isotropic polarizabilities and electronic spatial extents, together with frequency-derived correction values. By associating multiple geometries and calculated properties with each molecular fragment, QO2Mol supports the development and evaluation of atomistic machine-learning models as well as further analysis of geometry-dependent molecular properties.

\section{Methods}

\begin{figure}[tp]
\centering
\includegraphics[width=\linewidth]{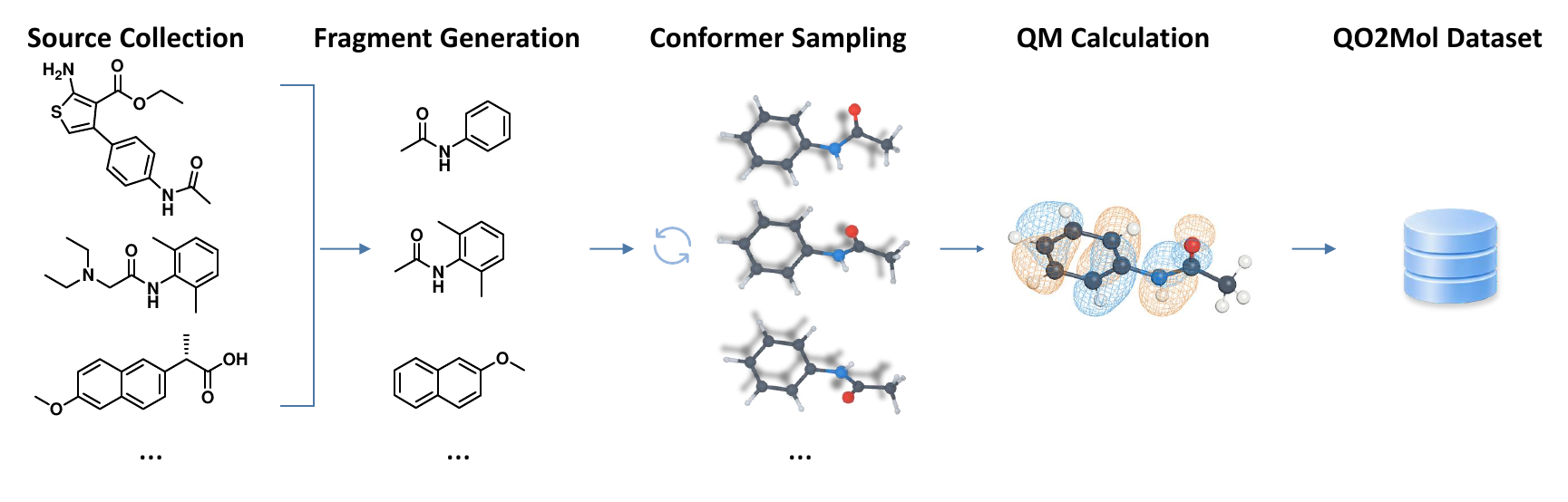}
\caption{Schematic overview of the QO2Mol data-generation workflow. Source compounds are processed by bond cleavage and capping to produce independent organic molecular fragments. Conformational sampling generates three-dimensional geometries, which are subjected to quantum-chemical calculations to obtain electronic energies, atomic forces and other calculated properties. The resulting records are assembled into the QO2Mol dataset.}
\label{fig:workflow}
\end{figure}

The data-generation workflow consists of molecular preparation, fragmentation, conformational sampling and quantum-chemical calculations as illustrated in Fig.~\ref{fig:workflow}.

\subsection{Source compounds and molecular preparation}

The initial molecular collection was obtained from ChEMBL version 33~\cite{Gaulton2012}, with molecular structures represented
as SMILES strings~\cite{Weininger1988}. These compounds were
processed to generate a library of independent organic molecular
fragments, as described in the following section. The retained
fragment collection was restricted to molecules composed of the following
ten elements: H, C, N, O, F, P, S, Cl, Br and I.

\subsection{Molecular fragmentation}

\begin{figure}[tbp]
\centering
\includegraphics[width=0.9\textwidth]{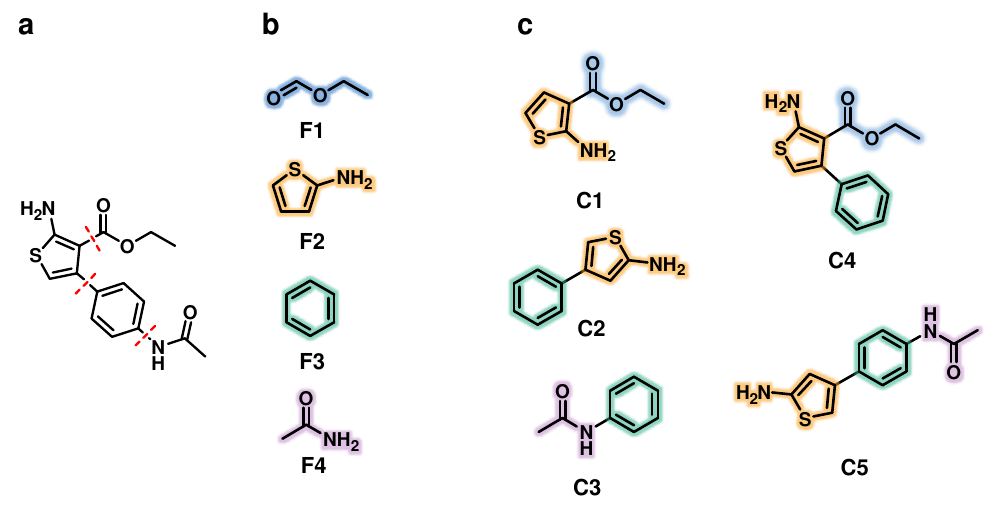}
\caption{Fragmentation of a representative source molecule. (\textbf{a}) Initial molecular structure with cleavage sites marked by red dashed lines. (\textbf{b}) Individual fragments F1--F4 obtained by cleaving the indicated bonds and capping exposed valences. (\textbf{c}) Larger connected fragments C1--C5 comprising different combinations of adjacent structural units from the original molecule. Colours identify corresponding substructures across the panels.}
\label{fig:fragment}
\end{figure}

Molecular fragments were generated by selective bond cleavage according to the procedure and settings provided in the accompanying code repository.
Cleaved boundaries were capped to satisfy atomic valences, yielding
independent molecular structures. Different selections of cleavage sites
within a source molecule can yield both smaller fragments and larger connected
substructures while preserving the original connectivity of the retained
portions, as illustrated in Fig.~\ref{fig:fragment}.
The resulting molecular structures were checked for consistency in molecular
connectivity, hydrogen-atom attachment, formal charge and stereochemical identity.

Successfully generated fragments were deduplicated using canonical isomeric
SMILES after removing atom-mapping labels and normalizing removable explicit
hydrogens. This representation preserves
distinctions in molecular connectivity, formal charge, isotopic composition
and specified stereochemistry. Duplicate occurrences were consolidated into
single molecular entries. Standard InChI and InChIKey identifiers were generated for molecular annotation. The resulting fragment collection was used for conformational
sampling and quantum-chemical calculations.

\subsection{Initial geometries and conformational sampling}
Initial three-dimensional geometries were generated from canonical SMILES and optimized at the B3LYP/def2-SVP level of theory. The unconstrained optimized geometries were screened by vibrational frequency analysis, and only those without imaginary frequencies were designated as equilibrium-reference structures. Rotatable bonds were identified from molecular connectivity, and the associated dihedral angles were systematically sampled in $30^\circ$ increments. Each sampled torsional configuration underwent constrained geometry optimization at the B3LYP/def2-SVP level, with the corresponding dihedral angles held fixed.

To sample additional non-equilibrium regions of the potential energy surface, selected geometries were subjected to bond-length perturbations of 0.05--0.08~\AA{}, with the magnitude depending on the bond type, and bond-angle perturbations of $5^\circ$--$10^\circ$. The resulting geometries were included in subsequent quantum-chemical calculations to obtain electronic energies and atomic forces.

The combination of constrained geometries and perturbed structures samples variations in energy away from local minima. The number of geometries generated for each molecule depends on its internal degrees of freedom and the sampling procedure.

\subsection{Quantum chemical calculations}
Electronic energies and Cartesian atomic forces were calculated for the sampled molecular geometries at the B3LYP/def2-SVP level of theory~\cite{Becke1993,Lee1988,Weigend2005}. All calculations were performed for the singlet electronic state, with the molecular charge specified for each fragment. The reported energies correspond to the electronic energies at the specified nuclear geometries, and the atomic forces are the negative Cartesian energy gradients.

Vibrational frequency calculations yielded zero-point
energy corrections and thermal corrections to energy,
enthalpy and Gibbs free energy at 298.15 K and 1 atm
under the harmonic-oscillator and ideal-gas approximations.
The electronic-structure calculations also provided
dipole moments, isotropic polarizabilities and
electronic spatial extents. The frequency-derived corrections were retained directly from the calculation outputs, without additional frequency cutoffs or quasi-harmonic corrections. Imaginary vibrational modes were omitted from the thermochemical analysis, whereas low positive frequencies were treated within the harmonic-oscillator approximation. 

\FloatBarrier

\section{Data Records}
The QO2Mol release is organized into files containing molecular-conformation records and their calculated properties. Each record describes a molecular fragment at a particular set of Cartesian coordinates. The \texttt{inchikey} field identifies a molecular structure\cite{Heller2015}, whereas \texttt{confid} identifies its conformation; both identifiers are needed when associating geometries with calculated property records. The README accompanying the Zenodo release documents the file names, formats, record counts and organization.

The released data fields are summarized in Table~\ref{tab:fields}. Atomic numbers, Cartesian coordinates and atomic forces share a consistent atom ordering.

\begin{table}[H]
\centering
\footnotesize
\setlength{\tabcolsep}{3.4pt}
\renewcommand{\arraystretch}{1.10}
\caption{Fields and calculated properties of QO2Mol molecular conformations. $N$ and $B$ denote the numbers of atoms and undirected bonds. A dash indicates that no physical unit applies.}
\label{tab:fields}
\begin{tabularx}{\linewidth}{@{}P{.23\linewidth}P{.155\linewidth}Y P{.09\linewidth}@{}}
\toprule
Property / field & Type / shape & Description & Unit \\
\midrule
inchikey & String & Molecular structure identifier shared by its conformations. & -- \\
confid & String & Identifier of a molecular conformation. & -- \\
atom\_count & Integer & Number of atoms ($N$). & -- \\
bond\_count & Integer & Number of undirected bonds ($B$). & -- \\
elements & $N$ integers & Atomic numbers in the stored atom order. & -- \\
coordinates & $N\times3$ & Cartesian atomic coordinates. & \AA{} \\
edge\_list & $2B\times2$ & Directed atom-index pairs; two entries per bond. & -- \\
edge\_attr & $2B$ values & Bond orders aligned with \texttt{edge\_list}: 1, single; 2, double; 3, triple. Aromatic bonds are represented in Kekul\'e form. & -- \\
net\_charge & Integer & Net molecular charge number, expressed in units of elementary charge. & $e$ \\
formal\_charge & $N$ integers & Atomic formal charge numbers in the stored atom order. & $e$ \\
energy & Scalar & Electronic energy at the recorded geometry. & eV \\
force & $N\times3$ & Cartesian atomic forces in the stored atom order. & eV/\AA{} \\
dipole & 3 values & Cartesian dipole-moment components in the input coordinate frame ($x$, $y$, $z$). & Debye \\
iso\_polar & Scalar & Isotropic electronic polarizability. & bohr$^3$ \\
ese & Scalar & Electronic spatial extent, $\langle R^2 \rangle$. & bohr$^2$ \\
zpe\_correction & Scalar & Zero-point energy correction from frequency analysis. & Ha \\
thermal\_corrections & Dictionary & Container for thermal corrections to energy, enthalpy and Gibbs free energy. & -- \\
Thermal correction to energy & Scalar & Key \texttt{energy} in \texttt{thermal\_corrections}; not total internal energy. & Ha \\
Thermal correction to enthalpy & Scalar & Key \texttt{enthalpy} in \texttt{thermal\_corrections}; not total enthalpy. & Ha \\
Thermal correction to Gibbs free energy & Scalar & Key \texttt{gibbs} in \texttt{thermal\_corrections}; not total Gibbs free energy. & Ha \\
\bottomrule
\end{tabularx}
\end{table}

\FloatBarrier

\section{Data Overview}
\begin{figure}[htbp]
\centering
\includegraphics[width=\linewidth]{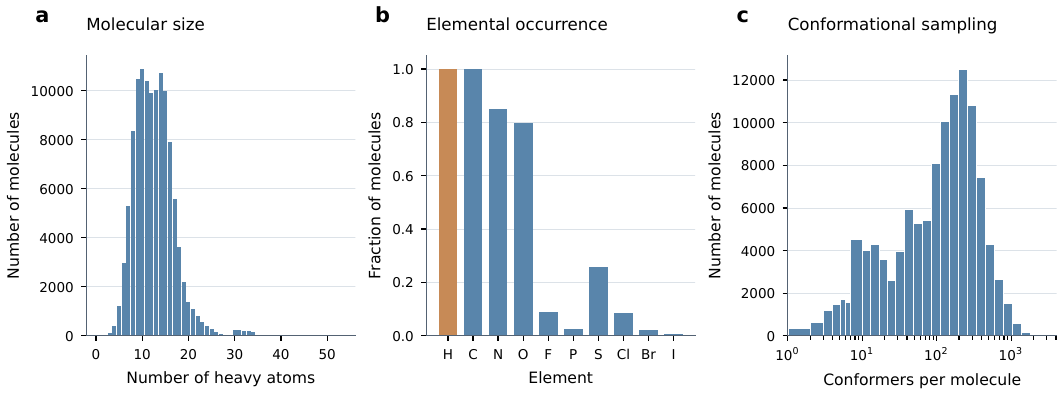}
\caption{Molecular composition and conformational sampling of QO2Mol. (\textbf{a}) Number of distinct molecules by heavy-atom count. (\textbf{b}) Fraction of distinct molecules containing each indicated element; a molecule may contribute to several element categories. (\textbf{c}) Distribution of the number of conformations per molecule, shown on a logarithmic horizontal axis. }
\label{fig:overview}
\end{figure}

QO2Mol comprises 116,127 organic molecular fragments represented by over 20 million three-dimensional conformations. The collection contains optimized and non-equilibrium geometries generated by torsional sampling and local structural perturbations. Figure~\ref{fig:overview} summarizes molecular size, elemental occurrence among distinct molecular fragments, and conformational sampling depth. Each molecule contributes once to the molecular-size and elemental-occurrence statistics, and the conformer-count distribution is evaluated at the molecular level.

\FloatBarrier

\section{Technical Validation}
The technical quality of QO2Mol was evaluated in terms of the reliability of the quantum-chemical calculations and the chemical plausibility of the resulting geometries. The released records were also checked for completeness and numerical validity. All required fields were present, and no missing values, NaNs or infinities were identified.

\subsection{Convergence and recovery of quantum-chemical calculations}
The quantum-chemical calculations were checked for SCF convergence. Calculations that did not converge on the first attempt were repeated at the same B3LYP/def2-SVP level of theory using an extended iteration limit of 256 cycles and an SCF procedure with a quadratically convergent fallback. A total of 298,987 calculations required recomputation. Of these, 298,905 completed successfully, whereas 82 did not terminate normally after the second attempt and were excluded from the dataset.

Unconstrained geometry optimizations employed the default gradient and displacement convergence criteria, including an RMS force threshold of $3.0\times10^{-4}$ Hartree/bohr. The vibrational frequency results of the unconstrained optimized geometries were screened for imaginary frequencies. Only geometries without imaginary frequencies were designated as equilibrium-reference structures; together with the optimization convergence criteria, this supports their assignment as local minima. These stationary-point criteria were not imposed on the deliberately sampled non-equilibrium geometries, including structures obtained by constrained torsional optimization and local geometric perturbations.

\subsection{Geometric validity and structural consistency}
The internal consistency of the released molecular records was verified by cross-checking atomic compositions, formal charges, net molecular charges and stored molecular connectivity. Atom counts were compared with the atomic-number arrays and the dimensions of the coordinate and force arrays, while bond counts were checked against the directed bond-index and bond-attribute representations. Molecular composition and connectivity were also checked across conformations sharing the same molecular identifier. Screening for molecular identity, fragmentation integrity, hydrogen-atom connectivity and stereochemical consistency resulted in the quarantine of 236,104 conformations. Of these, 13,324 were subsequently reinstated following verified corrections to molecular identity fields, without changes to atomic coordinates, molecular connectivity or quantum-chemical properties.

\begin{figure}[htbp]
\centering
\includegraphics[width=\linewidth]{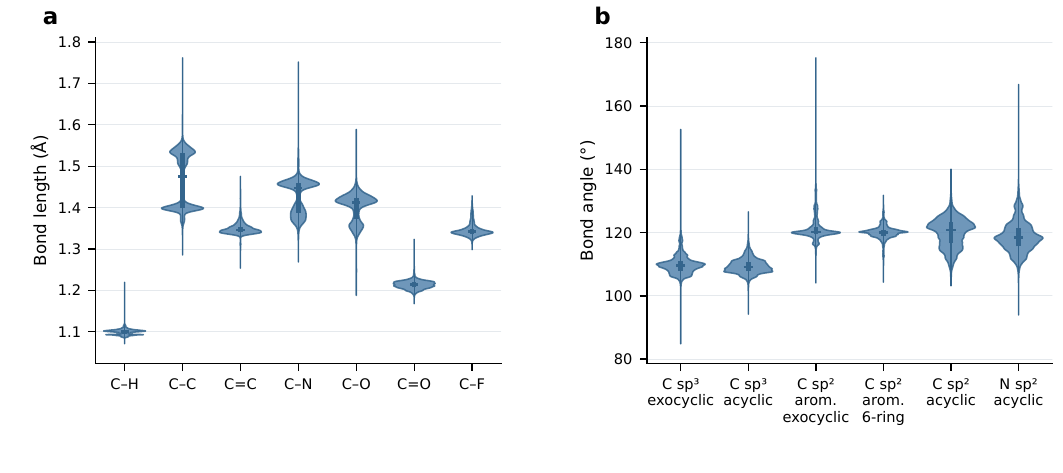}
\caption{Bond-length and bond-angle distributions of the equilibrium-reference structures in QO2Mol. Only retained unconstrained optimized geometries without imaginary frequencies are represented; deliberately generated non-equilibrium conformations are excluded. (\textbf{a}) Violin plots of bond lengths for selected bond types: C--H, C--C, C=C, C--N, C--O, C=O and C--F. (\textbf{b}) Violin plots of bond angles for selected local environments involving sp$^3$ carbon, sp$^2$ carbon and sp$^2$ nitrogen, as indicated by the horizontal-axis labels. Violin widths indicate the relative density of observed values within each category.}
\label{fig:validation}
\end{figure}

During dataset construction, molecular geometries were examined for inconsistencies between atomic coordinates, molecular connectivity and calculated properties. Optimized geometries were additionally screened for chemically implausible bond lengths with reference to the corresponding atom types and bond orders~\cite{Allen1987}. These checks led to the exclusion of 10,560 conformations with unresolved geometric or property-assignment inconsistencies. A further 67 conformations belonging to two molecular groups were excluded because the anomalous bond lengths could not be reconciled with a consistent molecular topology. 
The geometric characteristics of the retained equilibrium-reference structures are summarized in Fig.~\ref{fig:validation}.

\FloatBarrier

\section{Usage Notes}
The QO2Mol dataset is distributed in Python pickle format. Data preparation and processing scripts, together with examples of data loading and usage, are available in the QO2Mol GitHub repository at \url{https://github.com/kzhoa/QO2Mol}. When converting records into other representations, the original atom ordering should be preserved across atomic numbers, Cartesian coordinates and atomic forces. Coordinates, electronic energies and forces are reported in \AA{}, eV and eV/\AA{}, respectively. Python pickle files should be loaded only from trusted sources\cite{PythonPickle}.

The dataset contains both equilibrium-reference and non-equilibrium molecular geometries. Conformer counts reflect the sampling procedure rather than thermodynamic populations. Zero-point energy and thermal corrections are provided for all conformations. Their conventional thermochemical interpretation applies to the equilibrium-reference structures identified by unconstrained optimization and vibrational-frequency screening. For non-equilibrium geometries, these corrections should be treated as numerical calculation outputs.

Relative conformational energies should be compared only for structures with the same molecular composition and electronic state. For molecule-disjoint evaluations, all conformations belonging to the same molecular identity should be assigned to the same subset to avoid information leakage.

\section{Data Availability}
The QO2Mol dataset is deposited on Zenodo: \url{https://zenodo.org/records/23995922}.

\section{Code Availability}
The QO2Mol project repository, including publicly available processing and usage scripts, is hosted at \url{https://github.com/kzhoa/QO2Mol}.

Molecular processing and initial-geometry generation used RDKit version 2025.09.1~\cite{RDKit}. Quantum-chemical calculations were performed using Gaussian 16~\cite{Gaussian16}.

% \bibliographystyle{naturemag}
% \bibliography{references}

% --- end bbl

\end{document}